\documentclass[aps,prl,twocolumn,groupedaddress,showpacs,bibnotes]{revtex4}
\usepackage{graphicx,epsfig,subfigure,afterpage,bm}

\newcommand{\be}{\begin{equation}}
\newcommand{\ee}{\end{equation}}
\newcommand{\bea}{\begin{eqnarray}}
\newcommand{\eea}{\end{eqnarray}}
\newcommand{\gdot}{\dot{\gamma}}
\newcommand{\gdotbar}{\overline{\dot{\gamma}}}

\newcommand{\bw}{\begin{widetext}}
\newcommand{\ew}{\end{widetext}}

\newcommand{\tw}{t_{\rm w}}

\newcommand{\range}{\gdot_{\rm max}-\gdot_{\rm min}}

\begin{document}

\title{Age-dependent transient shear banding in soft glasses}
\author{Robyn L. Moorcroft$^{1}$, Michael E. Cates$^{2}$, and Suzanne M. Fielding$^{1}$}
\affiliation{$^{1}$Department of Physics, University of Durham, Science Laboratories, South Road, Durham. DH1 3LE, UK}
\affiliation{$^{2}$SUPA, School of Physics, University of Edinburgh, JCMB Kings Buildings, Mayfield Road, Edinburgh EH9 3JZ, UK}
\date{\today}
\begin{abstract} 

We study numerically the formation of long-lived transient shear bands
during shear startup within two models of soft glasses (a simple
fluidity model and an adapted `soft glassy rheology' model). The
degree and duration of banding depends strongly on the applied shear
rate, and on sample age before shearing. In both models the ultimate
steady flow state is homogeneous at all shear rates, consistent with
the underlying constitutive curve being monotonic. However,
particularly in the SGR case, the transient bands can be extremely
long lived. The banding instability is neither `purely viscous' nor
`purely elastic' in origin, but is closely associated with stress
overshoot in startup flow.

\end{abstract}
  \maketitle

Many soft materials show shear banding: the separation into layers of different shear rate under imposed flow. Examples include wormlike micelle solutions \cite{fluidbanding}; granular matter \cite{FenisteinNat03}; star polymers \cite{Rogers}; and `soft glasses' such as gels, pastes and emulsions \cite{OvarlezRheoAct09}. Because shear banding dramatically alters the stress response at fixed shear rate or vice versa, it is central to the rheological control of such materials, whose applications range from foodstuffs and pharmaceuticals to paints and well-bore fluids. For materials that are nonergodic at rest, unexpected complexity can arise when one band is not flowing and thus subject to aging \cite{Rogers,OvarlezRheoAct09,CoussotPRL02,softmatter}, while the other is continuously rejuvenated by flow \cite{MCT}. This interplay between glassy dynamics and shear banding brings together two major current areas of nonequilibrium physics research.

A steady mean shear rate is imposed by choosing time-independent wall
velocities in a rheometric device. In most reported cases, the
resulting shear bands then persist
indefinitely~\cite{Rogers,OvarlezRheoAct09,CoussotPRL02,Moller_lambdamodel,Goyon,Gibaud,Isabands,
VarnikPRL03_LJbanding}. In many instances, shear banding is
attributable to a nonmonotonic steady-state constitutive curve
$\Sigma(\dot\gamma)$, where $\Sigma$ is the shear stress and
$\dot\gamma$ the shear rate in the homogeneous material; regions with
$d\Sigma/d\dot\gamma <0$ are mechanically unstable
\cite{fluidbanding}. This is analogous to, but distinct from, a
similar instability in nonlinear elastic solids, where mechanical
instability likewise arises if the stress is a decreasing function of
strain ($d\Sigma/d\gamma <0$); see~\cite{MarrucciGrizzuti}.  For simplicity we refer
to these as `viscous' and `elastic' banding scenarios respectively.

In this Letter we use the case of soft glasses to explore
theoretically a distinctive, third scenario. This is where shear
banding arises on startup of steady shearing, before eventually
relaxing to an unbanded steady state. The resulting transient bands
can be extremely long-lived and so might well be mistaken for true
steady-state ones (see, e.g.~\cite{HuRecent}).  We show that such
bands can arise in systems {\em showing neither a viscous nor an
elastic instability}, such as a model combining essentially linear
elasticity with a near-trivial constitutive curve,
$\Sigma(\dot\gamma)= A+B\dot\gamma$.  Our results may thus prove
highly relevant (alongside other factors \cite{Wagner10,besseling}) to
a number of cases where apparently steady shear banding is seen, even
though the constitutive curve is predicted, by well founded theories,
to remain monotonic \cite{besseling,wangbands,AdamsOlmsted}. They are
also relevant to recent experiments where long-lived transient shear
bands are directly reported \cite{transientbands}.

We argue that transient banding behavior can arise generically in
systems where the stress response $\Sigma(t,\dot\gamma)$ to startup of
steady shear shows, as a function of time, a significant overshoot
before falling to the steady-state limiting value
$\Sigma(\infty,\dot\gamma)\equiv\Sigma(\dot\gamma)$. Indeed, the
simplest case is where the overshoot is created purely by nonlinear
elasticity without relaxation or plastic flow. In this case,
$\gamma(t) = \dot\gamma t$ is an elastic strain, so that any region
where $\partial_t\Sigma(t,\dot\gamma)<0$ implies the onset of the
elastic instability referred to already. In
reality however, soft glasses (in contrast to, e.g., elastomers \cite{MarrucciGrizzuti}) have
a very limited elastic deformation regime before they yield
irreversibly; the stress maximum in such cases is the result of a
contest between the growth of elastic stress and its decay by plastic
rearrangement towards the eventual steady-state limit. Accordingly
$\gamma(t)$ is not an elastic strain within the region where
$\partial_t\Sigma <0$, and there can be no direct mapping onto an
elastic banding scenario.

Below we show that transient shear banding does nonetheless arise,
whenever the stress overshoot becomes large, both in a fluidity model
of soft glasses (chosen for its tractability), and in a mesoscopic
model, adapted from that of \cite{JoR} which is known to capture well
some subtler physics of these systems. In soft glasses the overshoot
can be varied in height ({\em without} also varying
$\Sigma(\dot\gamma)$) purely by changing the age of the system
\cite{JoR}, making such materials ideal testing grounds for
experiments and theory on transient banding. Emerging as it does from
two independent models, transient shear banding should be a generic
feature in the rheology of well-aged soft glasses -- a fact not
appreciated previously. Moreover, stress overshoots often arise in
other types of soft matter, and in many of these, similar mechanisms
may be at work.

{\em Fluidity model:} Our first model is an empirical fluidity model, along the lines of \cite
{Picard,CoussotBonnPRL02_avalanche}, in which a single structural relaxation time $\tau$  regresses continuously towards a steady-state value determined by the local shear rate. A finite diffusivity for $\tau$ is also introduced, so as to prevent band interfaces from becoming infinitely sharp. This model shows transient banding in the overshoot region, with a strong dependence on system age, which sets the initial value of $\tau$ and thus the overshoot height. However the lifetime of the bands apparently remains modest, regardless of the system age before shearing begins, in contrast to the adapted SGR model considered later. 

To set up the fluidity model we decompose the total shear stress $\Sigma$ into a viscoelastic stress $\sigma$ arising from the glassy degrees of freedom and a purely viscous part:
$\Sigma=\sigma + \eta \gdot$.
We assume translational invariance in the $x$ (flow) and $z$ (vorticity) directions but allow nonuniformity to develop in the flow gradient direction, $y$. Force balance at zero Reynolds number (neglecting inertia) then requires $\Sigma$ to be independent of $y$. We suppose a Maxwell-type constitutive equation for the viscoelastic stress
\be
\partial_t\sigma=G\gdot-\sigma/\tau \label{eqn:sigma}
\ee
where $G$ is an elastic modulus and $\tau$ is the structural relaxation time (inverse fluidity) with its own dynamics:
\be
\partial_t\tau=1-\tau/\tilde\tau(\gdot)+l_o^2\partial^2_y \tau
\label{eqn:tau}
\ee
In the absence of flow this represents simple aging ($\tau = t$) but with flow present, aging is cut off at the inverse strain rate. (This is characteristic of strain-induced plasticity \cite{MCT}.)
Here $l_o$ is a mesoscopic length (which could depend on $\dot\gamma$
without affecting our presentation) describing the tendency for the relaxation time of a mesoscopic region to equalise with those of its neighbors. 

We choose the steady-state relaxation time as
$
\tilde\tau=\tau_0+\lambda/|\gdot|$,
so that in steady state $\Sigma = \sigma +\eta\dot\gamma$ with $\sigma = G\dot\gamma\tilde\tau = G\lambda + G\tau_0\dot\gamma$, has a yield stress beyond which it is trivially monotonic (a Bingham fluid).  
We consider flow between infinite flat parallel plates at $y=0,
L_y$, and rescale strain, stress, time and length so that $\lambda = G=\tau_0=L_y=1$.  The steady-state constitutive curve is then
\be
\Sigma(\dot\gamma) = 1 + (1+\eta)\dot\gamma \label{eqn:cc}
\ee
where for simplicity we  now set $\eta=0.05$.

We study a shear startup protocol defined as
follows. First imagine preparing the sample by a deep quench at time $t=0$, which we
assume results in a fully rejuvenated initial state with $\tau(y,t=0)=1,
\sigma(y,t=0)=0$ across the whole sample. Next we allow the sample to age
at rest (so $\dot\tau=1$) until a time $\tw$, before setting
the upper plate moving along $x$ with constant speed $\gdotbar
L_y$. This defines the average imposed shear rate $\gdotbar = \int_0^1\gdot(y,t)dy$.

\begin{figure}[tb]
\includegraphics[width=8.0cm]
{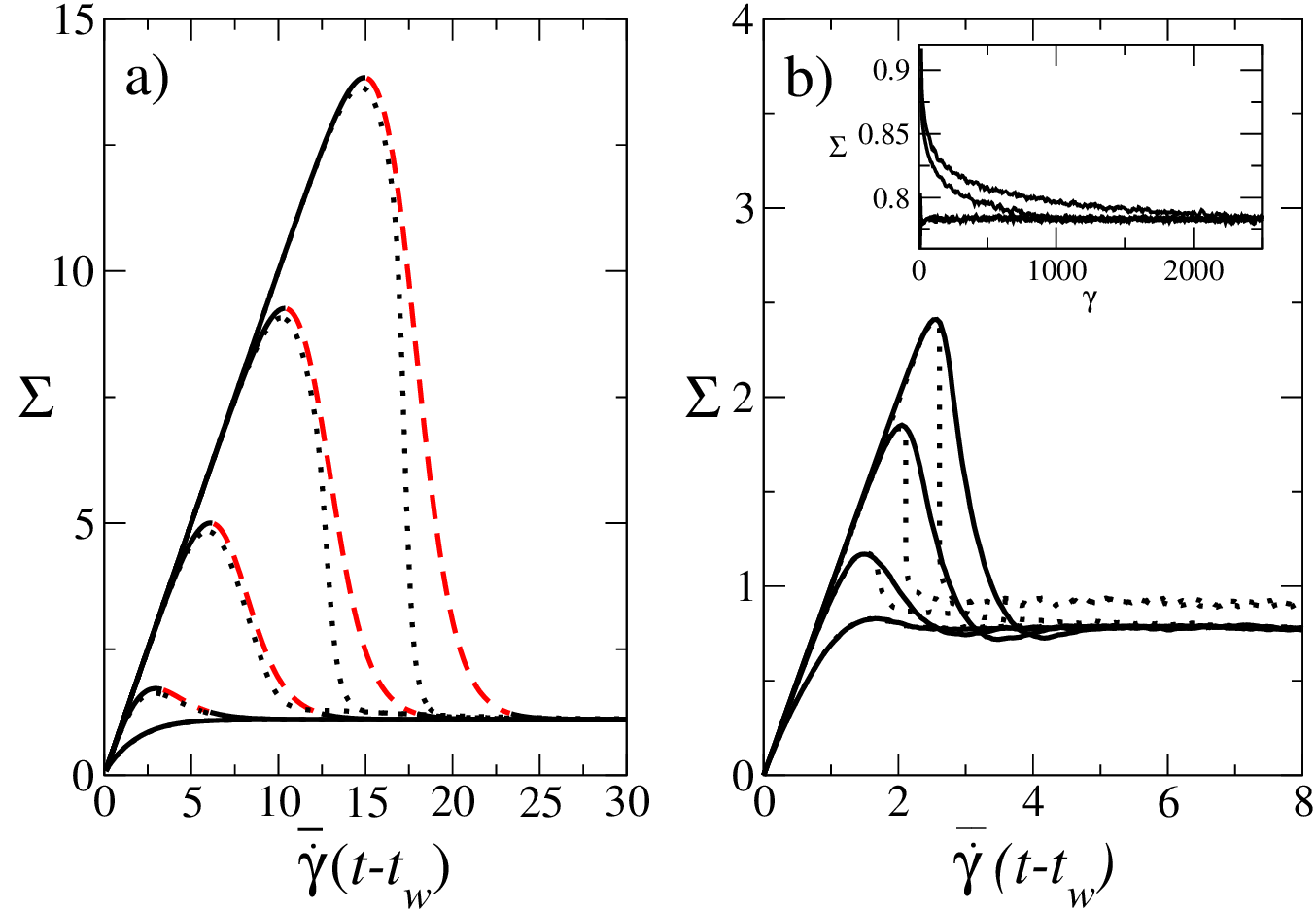}
  \caption{ (a) Homogeneous startup transient in fluidity model at
  applied shear rate $\gdotbar=0.1$ for waiting times
  $\tw=10^0,10^2,10^4,10^6,10^8$ ($\tw$ increasing with increasing
  size of overshoot) showing stable region (solid), unstable
  (dashed). Dotted curve: actual startup curve arising from
  inhomogeneous flow with $l_0=0.01$ and $\delta=0.01$. (b)
  Counterpart in modified SGR model for $x=0.3, \gdotbar=0.1,
  w=0.05,n=50,m=300$, and waiting times
  $\tw=10^0,10^2,10^4,10^6$. Solid lines: banding disallowed. Dotted:
  banding allowed. Inset: approach to steady state at later times in
  the banding case.}
\label{fig:overshoots}
\end{figure}

Transient stress curves $\Sigma(t,\dot\gamma)$ for shear startup in a homogeneous system are shown for several different $\tw$ in Fig.~\ref{fig:overshoots}a. There is an overshoot that depends strongly on the age $\tw$ of the
sample; one may show that it occurs at a strain
$\gamma_o=\gdot(t-\tw)$ given by $\gamma_o\exp(\gamma_o)=\gdot\tw$.
To within a logarithmic correction this gives
$\gamma_o=\log(\gdot\tw)$, which we use for convenience below in
preference to the full implicit expression. The response of the sample
prior to (but not beyond) the stress maximum is almost elastic, so the
peak stress obeys $\Sigma_o\approx\gamma_o$.

\begin{figure}[tb]
  \includegraphics[width=8.0cm]{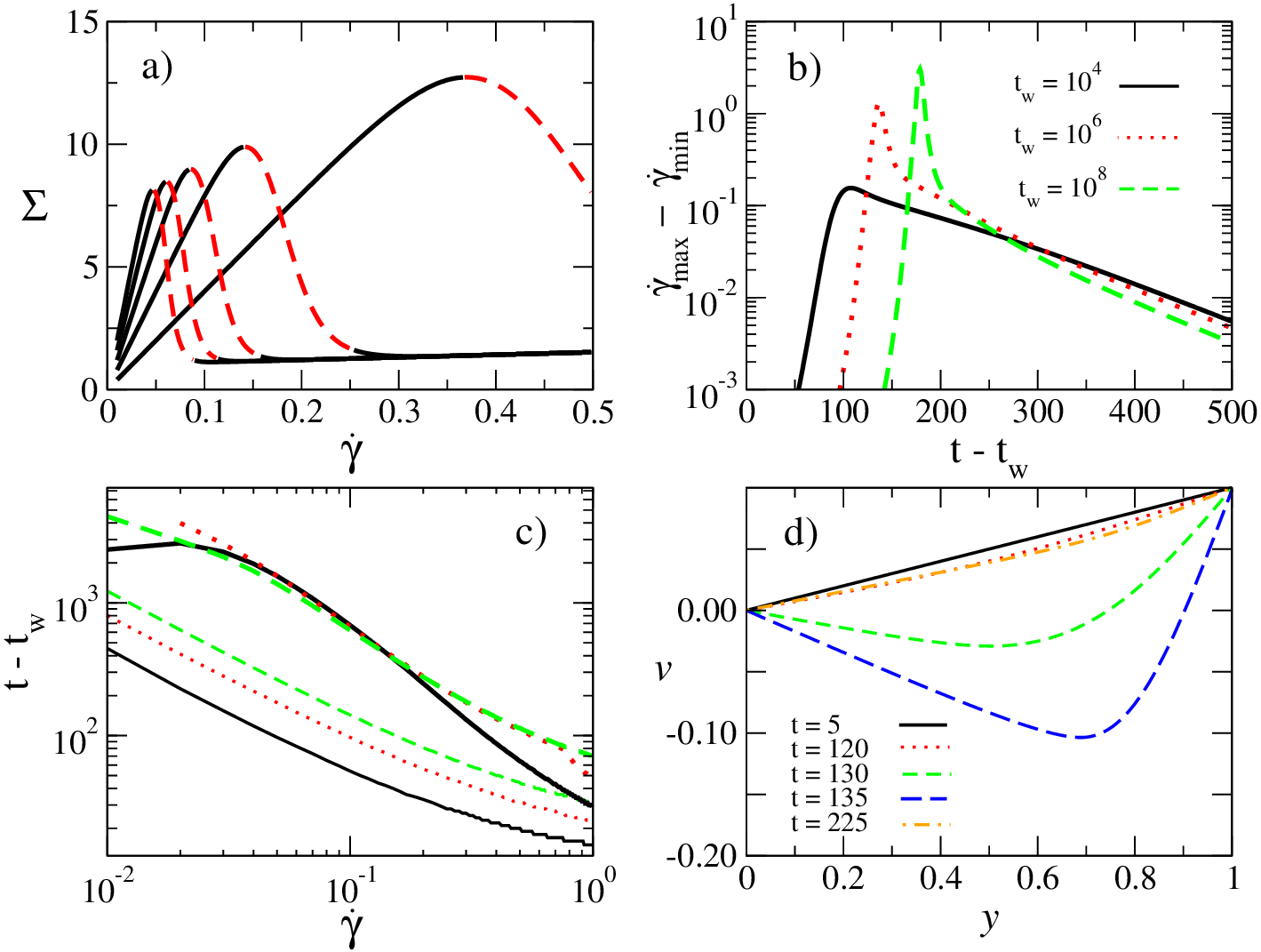}
  \caption{(a) For chosen $t_w = 10^6$, plots of the instantaneous
  homogeneous constitutive curves during start up for $t-t_w = 40, 80,
  120, 160, 200$ with regions of instability dashed.  (b) Strength of
  banding ($\dot\gamma_{\rm max}-\dot\gamma_{\rm min}$) against
  $t-t_w$ for $\gdotbar = 0.1$, and $t_w = 10^4,10^6,10^8$.  (c) For
  the same $\gdotbar=0.1$, and $t_w = 10^4$ (solid), $10^6$ (dotted),
  $10^8$ (dashed), plots in the $t-t_w,\dot\gamma$ plane showing the
  upper (thick lines) and lower (thin lines) boundary curves of the
  window in which banding is significant (threshold set by
  $(\range)/\gdotbar>0.01$). (d) Snapshots of banded profiles for
  $\dot\gamma = 0.1$, $t_w = 10^6$, and $t-t_w = 5,120,130,135,225$
  showing onset, development and eventual collapse of the
  bands. $l_0=0.01$, $\delta=0.01$ in  (b-d).}
\label{fig:TWO}
\end{figure}

To gain intuition for the consequences of the overshoot, consider now
a thought experiment in which one tracks this homogeneous stress
transient for several identical sample replicas, each subject to a
different shear rate. At any fixed time interval $t-\tw$ this defines
an `instantaneous constitutive curve' $\tilde\Sigma(\gdot)$, with the
tilde denoting a dependence on $t-\tw$ and $\tw$, suppressed for
clarity of notation. As shown in Fig.~\ref{fig:TWO}(a), each such
curve is non-monotonic over a typical shear rate window $\gdot <
\gamma_0/(t-\tw)$, with monotonicity being restored at higher shear
rates. This can be understood as follows. At any fixed time interval
$t-\tw$, those replicas at high shear rate $\gdot(t-\tw)>\gamma_0$ are
expected to have reached steady state, with stresses obeying the
constitutive curve, Eq.(\ref{eqn:cc}). In contrast those for which
$\gdot(t-\tw)<\gamma_0$ will still be on the elastic branch of the
stress transient, with a stress $\sigma=\gdot(t-\tw) \gg
\Sigma(\dot\gamma)$.  With this transient non-monotonicity in mind, we
performed an instantaneous linear stability analysis about the
evolving homogeneous flow, determining for each $t-\tw$ an
instantaneous eigenvalue whose positivity indicates instability to the
onset of banding. The unstable windows are shown by dashed lines in
Fig.~\ref{fig:overshoots}a, and correspond broadly to the regions of
negative slope in $\tilde\Sigma(\gdot)$. However, they do not do so
exactly; this cannot be viewed solely as a `viscous' instability of
the instantaneous constitutive curve. Nor does the window exactly
correspond to that of negative slope in $\Sigma(t,\gdot)$, as it would
for an `elastic' instability. Although clearly associated with a large
stress overshoot, the banding scenario found here is thus distinct
from either the viscous or the elastic one -- a view reinforced by the
linearity of the step strain response in (\ref{eqn:sigma}), and the
trivial monotonicity in (\ref{eqn:cc}), for the underlying model.

We now turn to study the full heterogeneous dynamics of the model in
this shear startup protocol, allowing spatial variations in the flow
gradient direction $y$. In each run we add a small perturbation
$\sigma(y,t=0)=\delta \cos(\pi y)$ where $\delta\ll 1$, in order to
trigger any banding.  In each
run we tracked, as a function of shearing time $t-\tw$, the degree of
shear banding in the sample, as measured at any instant by the
difference $\range$ between the maximum and minimum shear rates
present in the cell, as shown in Fig.~\ref{fig:TWO}(b) for several
different waiting times $\tw$. Regions of high $\range$ broadly match
up with the regions of negative slope in Fig.~\ref{fig:TWO}(a),
consistent with the idea that this instability in the instantaneous
constitutive curve indeed triggers banding. Likewise in
Fig.~\ref{fig:TWO}(c) we delineate the temporal windows, for various
$\tw$ and as a function of shear rate $\gdotbar$, within which banding
is significant, as defined by a criterion $(\range)/\gdotbar>0.01$.  The
shear stress that arises in these transiently banded flows is of
course different from the predictions of the homogeneous model; both
are shown in Fig.~\ref{fig:overshoots}a, as a function of $t-\tw$, at
a fixed shear rate $\gdotbar=0.1$ for various waiting times.
Snapshots of the transient bands for one particular startup are shown
in Fig.~\ref{fig:TWO} (d). Note that the minimum shear rate, which
arises within the low shear-rate band, can be negative
during transient banding. This is because the low shear-rate band is
well-aged, has been subjected only to a weak flow, and therefore
remains essentially an elastic slab. However the stress $\Sigma$ on
this slab decreases post-overshoot: like any elastic solid being
unloaded, it shears backwards.

{\em Adapted SGR model:} Our second model is a spatially resolved
adaptation of the SGR model \cite{JoR,softmatter} in which a spectrum
of jump rates describes hopping over strain-modulated local
rearrangement barriers at an effective noise temperature $x = T/T_g$;
elastic strain builds up and then is released in these plastic jump
events. Following Ref.~\cite{softmatter}, where full details can be
found, numerically we take $j=1...m$ SGR elements on each of $i=1...n$
streamlines, corresponding to $y=0...1$, with periodic boundary
conditions. The stress on streamline $i$ is $\sigma_i = (k/m)\sum_j
\ell_{ij}$, with elastic constant $k=1$. A waiting time Monte Carlo
algorithm is used to choose stochastically the next element to
jump. Supposing the jump occurs at element $ij$ when its local strain
is $\ell = l$, force balance is then imposed by updating all elements
on the same streamline as $\ell\to \ell + l/m$, and further updating
all elements throughout the system as $\ell \to \ell -l/mn$. In
contrast to Ref.~\cite{softmatter}, here we take the noise temperature
$x$ as constant, but include instead a stochastic jump-induced
straining of elements on neighboring streamlines after each jump
event: further adjusting the strain of three randomly chosen elements
on each adjacent streamline $i\pm 1$ by $lw(-1,+2,-1)$. This creates a
diffusive coupling of the dynamics on different streamlines (different
$y$), analogous to that in Eq.(\ref{eqn:tau}), with a strength set by
$w$. It also mildly alters the constitutive curve from that of pure
SGR, without losing monotonicity.

The evolution of stress with time in startup flow is shown in
Fig.~\ref{fig:overshoots}b, both for the case where homogeneous flow
is imposed (solid lines), and for the full dynamics allowing banding
(dotted lines). The trends are quite similar to those from the scalar
model: for older samples, the dotted line departs from the solid one
just after the overshoot. The initial effect of transient banding, as
before, is always to decrease the stress to values below that of the
homogeneous system at the same time point.

\begin{figure}[tb]
  \includegraphics[width=6.25cm]{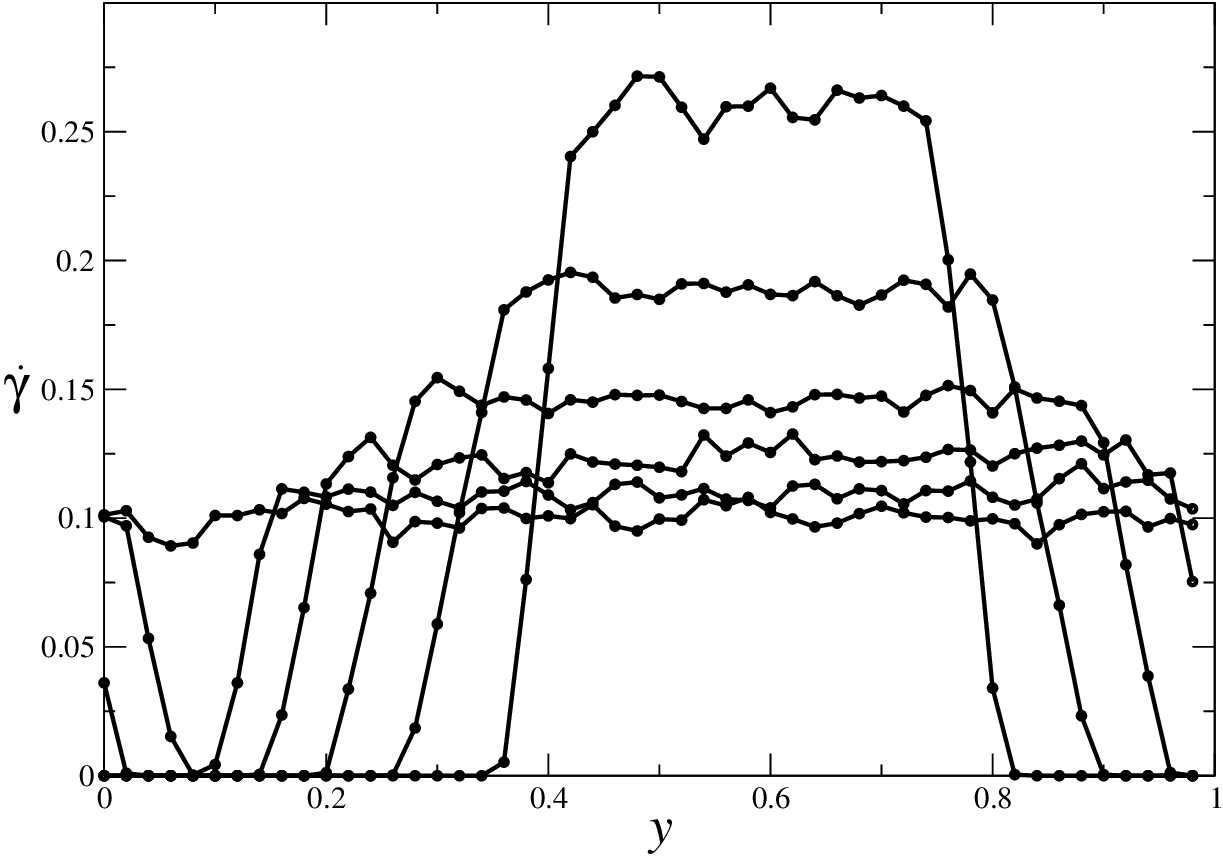}
  \caption{Shear rate profiles corresponding to slowly decaying stress
  signal at $\tw=10^6$ in Fig.~\ref{fig:overshoots}b averaged over
  time windows
  $250-500;500-1000;1000-1500;1500-2000;2000-2500;2500-3000$ (top to
  bottom at $y=0.5$).}
\label{fig:profiles}
\end{figure}

There is however a new feature in the SGR model, not seen in the
fluidity model. For the oldest samples ($t_w \sim 10^6$, measured in
units of the microscopic attempt rate for jumps) the time scale for
the stress signal to decay to the limiting value (corresponding to
homogeneous flow) is inordinately long: indeed, this decay can require
strains $\gdotbar t$ of order thousands, as opposed to the order-unity
values seen in the fluidity model. Since strain rejuvenates the
material, this is possible only because, for very old samples, the
strain rate in the low-shear band remains extremely small compared to
$\gdotbar$.
This behavior, clearly visible in the banding profiles presented in Fig.~\ref{fig:profiles}, is consistent with having an age-dependent static yield stress that can lie well above the homogeneous steady state stress $\Sigma(\gdotbar)$. In principle the banded state might then persist indefinitely, but in our model the low shear band seemingly is eroded slowly by the spreading of the fast band, perhaps as a result of the diffusive nonlocality in the jump dynamics. 

{\em Discussion:} Aside from its longer-lived transients, the adapted SGR model shows broadly similar phenomenology to the fluidity model, supporting our contention that the transient shear-banding scenario reported here is generic for soft glasses. As previously stated, the soft glass case is one where the stress overshoot stems from a competition between (essentially linear) elasticity and plastic relaxation: there is no facile mapping onto either an elastic instability ($d\Sigma/d\gamma <0$) or a viscous one ($d\Sigma/d\dot\gamma <0$). Despite the intermediate character of the instability towards transient bands, we showed for the fluidity model that their onset is closely correlated with the occurrence a negative slope on the {\em instantaneous} stress versus strain rate curve ($\partial\Sigma/\partial\dot\gamma|_t<0$). However, linearization about the time-dependent solution for a homogeneous flow showed this to be a qualitative rather than an exact correspondence.

The scenario of transient bands we have developed for soft glasses may interact in a complicated manner with various mechanisms previously presented to explain steady-state shear bands in the same class of materials \cite{besseling,softmatter}. However, in some cases bands of long but finite duration were already observed in startup flows \cite{transientbands} and the physics we have explored in this letter may be enough, on its own, to explain some of these. Even if not, the presence of a generic connection between transient banding and stress overshoots should not be overlooked in future work on these materials. 

Stress overshoots in startup flows are often also seen in other classes of viscoelastic soft matter at high enough shear rates, including non-aging systems such as entangled polymers. Indeed, some studies of polymeric materials and models have started to explore the connection between this and transient banding \cite{wangbands,AdamsOlmsted}. Nonetheless, we hope that further experimental and theoretical work on soft glasses will help elucidate this connection in more general terms. Arguably these represent ideal materials with which to explore the problem, because the size of the overshoot can be varied without any change to the final steady state of the system. In all systems both the final state and the overshoot depend on the imposed flow rate and on sample composition, temperature etc.; but in glasses one can also vary the system age $t_w$ which affects only the overshoot, and not the steady state.

{\it Acknowledgments---} We thank Tom McLeish, Ron Larson, Peter Olmsted and Rut Besseling for discussions. Work funded in part by EPSRC EP/E5336X/2 and EP/E030173. MEC is funded by the Royal Society.

\if{


\bibitem{expbanding}T. G. Mason
{\it et al.}, 
J. Col. Int Sci. {\bf 179}, 439 (1996);
F. Rouyer {\it et al.}, Phys. Rev E {\bf 67}, 021405 (2003); L.
Becu {\it et al.}, Phys. Rev. Lett {\bf 96}, 138302 (2006); G.
Ovarlez {\it et al.}, Phys. Rev. E {\bf 78} 036307 (2008); G.
Katgert {\it et al.}, Europhys. Lett. {\bf 90}, 54002 (2010).

\bibitem{ShiFalkPRL07_STZbanding}Y. Shi {\it  et al.}, Phys. Rev. Lett.  {\bf 98}, 185505 (2007).

\bibitem{BocquetPRL09_KEPmodel}L. Bocquet {\it et al.}, Phys. Rev. Lett. {\bf 103}, 036001 (2009).

\bibitem{microlength}R. Yamamoto, A. Onuki, Phys. Rev. E {\bf 58}, 3515 (1998); G. Picard {\it  et al.}, Phys. Rev. E {\bf 71}, 010501 (2005); A. Lema\^{i}tre, C. Caroli, Phys. Rev. Lett. {\bf 103}, 065501
(2009).

\bibliographystyle{prsty}
\bibliography{ackerson,actin,articles,banding,barham,Berret,berrportdecruppe,berthier,books,bray,callaghan,cates,chandcits,cook,crystal,crystal_theory,Decruppe,dhont,dnatheory,elasticTurbulence,extra,fielding,fischer,Fischer,flowcryst,fredrickson,gelbart,goveas,graham,Groisman,head,hebraud,helfrich,HinchRallison,hsiao,Kadoma,larson,LCtheory,leal,lerouge,lerougeDecruppe,lifshitz,line_tension,maffettone,malkus,master,mccoy,membs,Mexican,new,noirez,notes,olmsted,olmsted09,onions,otherRelated,phanThien,phd1,phd,pine,PineHu,pomeau,poon,pratt,psolutions,ramaswamy,recent,rheochaos,rheofolks,ryan,salmon,SalmonManneville,savedrecs.txt,schoot,semenov,sgrband,shaqfeh,sood,sriram,stein,sureshkumar,taylor,vansaarloos,vorticity,Wang,weitz,wilson,worms2,worms3,worms,yuan,zubarev}
\fi

\end{document}